\documentclass[preprintnumbers,amsmath,amssymbm,prd]{revtex4}
\usepackage{epsfig}
\usepackage{graphicx}

\begin{document}
\title{Quasinormal resonances of a charged scalar field in a charged Reissner-Nordstr\"om black
hole spacetime: A WKB analysis}
\author{Shahar Hod}
\affiliation{The Ruppin Academic Center, Emeq Hefer 40250, Israel}
\affiliation{ }
\affiliation{The Hadassah Institute, Jerusalem 91010, Israel}
\date{\today}

\begin{abstract}
\ \ \ The fundamental quasinormal resonances of charged
Reissner-Nordstr\"om black holes due to charged scalar perturbations
are derived {\it analytically}. In the WKB regime, $qQ\gg\hbar$, we
obtain a simple expression for the fundamental quasinormal
resonances: $\omega=qQ/r_+-i2\pi T_{\text{BH}}(n+{1 \over 2})$,
where $T_{\text{BH}}$ and $Q$ are the temperature and electric
charge of the black hole and $q$ is the electric charge of the
field. Remarkably, our results show that the relaxation dynamics of
a perturbed Reissner-Nordstr\"om black hole may saturate the
recently proposed universal relaxation bound.
\end{abstract}
\bigskip
\maketitle

\section{Introduction}

The uniqueness theorems \cite{un1,un2,un3} imply that the metric
outside a newly born black hole should relax into a Kerr-Newman
spacetime, characterized solely by the black-hole mass, charge, and
angular momentum. This implies that perturbation fields left outside
the newly born black hole would either be radiated away to infinity
or swallowed by the black hole. The relaxation phase in the dynamics
of perturbed black holes is characterized by `quasinormal ringing',
damped oscillations with a discrete spectrum (see e.g.
\cite{Nollert1,Ber1} for detailed reviews). These characteristic
oscillations are then followed by late-time decaying tails
\cite{Tails1,Tails2}.

The black hole quasinormal modes (QNMs) correspond to solutions of
the perturbation equations with the physical boundary conditions of
purely outgoing waves at spatial infinity and purely ingoing waves
crossing the event horizon \cite{Detw}. Such boundary conditions
single out a discrete set of complex black-hole resonances
$\{\omega_n\}$ (assuming a time dependence of the form $e^{-i\omega
t}$). In analogy with standard scattering theory, the QNMs can be
regarded as the scattering resonances of the black-hole spacetime.
It turns out that there exist an infinite number of quasinormal
modes, characterizing oscillations with decreasing relaxation times
(increasing imaginary part) \cite{Leaver,Hodsc,Kesh1}.

Quasinormal resonances are expected to be excited by a variety of
astrophysical processes involving black holes. Being the
characteristic 'sound' of the black hole itself, these free
oscillations are of great importance from the theoretical
\cite{Hodsc,Gary} and astrophysical point of view
\cite{Nollert1,Ber1}. They allow a direct way of identifying the
spacetime parameters. This has motivated a flurry of research during
the last four decades aiming to compute the resonance spectrum of
various types of black holes \cite{Nollert1,Ber1}.

In this work we consider the nearly spherical gravitational collapse
of a charged scalar field to form a charged Reissner-Nordstr\"om
black hole. The focus would be on the relaxation phase of the
charged perturbation fields which were left outside the newly born
black hole. In particular, we shall determine the fundamental
resonant frequencies which characterize the relaxation dynamics of
the charged fields in the charged black-hole spacetime. As we shall
show below, the spectrum of charged quasinormal resonances can be
studied {\it analytically} in the $qQ\gg \hbar$ regime (see details
below).

\section{Description of the system}

The physical system we consider consists of a charged scalar field
coupled to a charged Reissner-Nordstr\"om (RN) black hole. The
black-hole spacetime is described by the line element
\begin{equation}\label{Eq1}
ds^2=-f(r)dt^2+{1\over{f(r)}}dr^2+r^2(d\theta^2+\sin^2\theta
d\phi^2)\ ,
\end{equation}
where
\begin{equation}\label{Eq2}
f(r)\equiv 1-{{2M}\over{r}}+{{Q^2}\over{r^2}}\  .
\end{equation}
Here $M$ and $Q$ are the mass and electric charge of the black hole
respectively, and $r$ is the Schwarzschild areal coordinate. (We use
natural units in which $G=c=\hbar=1$.)

The dynamics of a charged scalar field $\Psi$ in the RN spacetime is
governed by the Klein-Gordon equation \cite{Hod3}
\begin{equation}\label{Eq3}
[(\nabla^\nu-iqA^\nu)(\nabla_{\nu}-iqA_{\nu})]\Psi=0\  ,
\end{equation}
where $A_{\nu}=-\delta_{\nu}^{0}{Q/r}$ is the electromagnetic
potential of the black hole and $q$ is the charge coupling constant
of the field. [Note that $q$ stands for $q/\hbar$ and thus it has
the dimensions of $($length$)^{-1}$.] One may decompose the field as
\begin{equation}\label{Eq4}
\Psi_{lm}(t,r,\theta,\phi)=e^{im\phi}S_{lm}(\theta)R_{lm}(r)e^{-i\omega
t}\ ,
\end{equation}
where $\omega$ is the conserved frequency of the mode, $l$ is the
spherical harmonic index, and $m$ is the azimuthal harmonic index
with $-l\leq m\leq l$. (We shall henceforth omit the indices $l$ and
$m$ for brevity.) The sign of $\omega_I$ determines whether the
solution is decaying $(\omega_I<0)$ or growing $(\omega_I>0)$ in
time.

With the decomposition (\ref{Eq4}), $R$ and $S$ obey radial and
angular equations both of confluent Heun type coupled by a
separation constant $K_l=l(l+1)$
\cite{Heun,Flam,Fiz1,Fiz2,Fiz3,HodSO}. The radial Klein-Gordon
equation is given by \cite{HodCQG2}
\begin{equation}\label{Eq5}
r^2f(r){{d^2R}\over{dr^2}}+(2r-2M){{dR}\over{dr}}-\Big[K_l-{{r^4}\over{\Delta}}(\omega-{{qQ}\over{r}})^2\Big]R=0\
.
\end{equation}
It is convenient to define a new ``tortoise" radial coordinate $y$
by
\begin{equation}\label{Eq6}
dy\equiv dr/f(r)\  ,
\end{equation}
in terms of which the radial equation (\ref{Eq5}) can be written in
the form of a Schr\"odinger-like wave equation
\begin{equation}\label{Eq7}
{{d^2\psi}\over{dy^2}}-V\psi=0\  ,
\end{equation}
where $\psi\equiv rR$. The effective potential in (\ref{Eq7}) is
given by
\begin{equation}\label{Eq8}
V=V(r;\omega,M,Q,q,l)=-\Big(\omega-{{qQ}\over{r}}\Big)^2+{{f(r)H(r)}\over{r^2}}\
,
\end{equation}
where the dimensionless function $H$ is defined by
\begin{equation}\label{Eq9}
H(r;M,Q,l)\equiv l(l+1)+{{2M}\over{r}}-{{2Q^2}\over{r^2}}\ .
\end{equation}
Note that $H\leq l(l+1)+1$ in the black-hole exterior.

We are interested in solutions of the radial equation (\ref{Eq7})
with the physical boundary conditions of purely outgoing waves at
spatial infinity and purely ingoing waves at the black-hole horizon.
That is,
\begin{equation}\label{Eq10}
%\label{eq:boundary_conditions}
\psi \sim
\begin{cases}
y^{-iqQ}e^{i\omega y} & \text{ as }
r\rightarrow\infty\ \ (y\rightarrow \infty)\ ; \\
e^{-i (\omega-qQ/r_+)y} & \text{ as } r\rightarrow r_+\ \
(y\rightarrow -\infty)\ .
\end{cases}
\end{equation}
These boundary conditions single out a discrete set of resonances
$\{\omega_n\}$ which correspond to the quasinormal resonances of the
field \cite{Dolan}.

\section{The quasinormal resonances}

We shall now show that the fundamental quasinormal resonances of
charged Reissner-Nordstr\"om black holes due to charged scalar
perturbations can be studied analytically in the regime
\begin{equation}\label{Eq11}
qQ\gg l+1\  .
\end{equation}
In this regime the effective potential (\ref{Eq8}) is in the form of
a potential barrier whose maximum lies in the vicinity of the
black-hole horizon. To see this, it is convenient to define new
dimensionless variables
\begin{equation}\label{Eq12}
x\equiv 1-{{r_+}\over{r}}\ \ \ ; \ \ \ \varpi\equiv {{\omega
r_+}\over{qQ}}-1\  ,
\end{equation}
in terms of which the effective potential (\ref{Eq8}) becomes
\begin{equation}\label{Eq13}
V(x;\varpi)=-\Big({{qQ}\over{r_+}}\Big)^2(x+\varpi)^2+{{H(r_+)(r_+-r_-)}\over{r^3_+}}x
[1+O({{xr_+}\over{r_+-r_-}})]\  .
\end{equation}
Here $r_{\pm}=M\pm(M^2-Q^2)^{1/2}$ are the black-hole (event and
inner) horizons. From (\ref{Eq13}) one finds
\begin{equation}\label{Eq14}
x_0+\varpi={{H(r_+)}\over{2q^2Q^2}}{{r_+-r_-}\over{r_+}}\ll 1\
\end{equation}
for the location $x_0$ of the peak of the effective potential. The
last inequality in (\ref{Eq14}) follows from (\ref{Eq9}) and
(\ref{Eq11}) which imply $H(r_+)\leq l(l+1)+1\leq (l+1)^2\ll
(qQ)^2$.

WKB methods \cite{WKB1,WKB2,WKB3,Will,CarC} provide an accurate
approximation for QNM frequencies in the eikonal limit $\omega
r_+\gg 1$. As emphasized in \cite{Will}, the WKB technique is based
on the similarity between the black hole perturbation equation
(\ref{Eq7}) and the Schr\"odinger equation for the case of a
potential barrier \cite{Will}. Following the analysis of
\cite{WKB1,WKB2,WKB3,Will}, one realizes that the quasinormal mode
condition for $\omega$ follows from the equations which connect the
amplitudes of the waves on both sides of the potential barrier,
together with the physical boundary conditions of outgoing waves at
both $y\to\pm\infty$. In particular, the QNM condition in the
eikonal limit is given by \cite{WKB1,WKB2,WKB3,Will,CarC}:
\begin{equation}\label{Eq15}
{{V(y_0)}\over{\sqrt{-2V^{(2)}(y_0)}}}=i(n+{1/2})\  ,
\end{equation}
where $V^{(2)}\equiv d^2V/dy^2$ and Eq. (\ref{Eq15}) is evaluated at
the extremum $y=y_0$ of $V(y)$.

The WKB condition (\ref{Eq15}) can be solved {\it analytically} in
the $qQ\gg l+1$ regime. Substituting Eqs. (\ref{Eq8}) and
(\ref{Eq14}) into Eq. (\ref{Eq15}), one finds
\begin{equation}\label{Eq16}
\varpi_R={{H(r_+)(r_+-r_-)}\over{4q^2Q^2r_+}}\cdot
{{H^2(r_+)+8q^2Q^2(n+1/2)^2}\over{H^2(r_+)+4q^2Q^2(n+1/2)^2}}\
\end{equation}
and
\begin{equation}\label{Eq17}
\varpi_I=-i{{H(r_+)(r_+-r_-)}\over{4q^2Q^2r_+}}\cdot
{{2qQH(r_+)(n+1/2)}\over{H^2(r_+)+4q^2Q^2(n+1/2)^2}}\  .
\end{equation}

It should be emphasized that the QNM condition, Eq. (\ref{Eq15}), is
valid provided higher-order correction terms which appear in the WKB
expansion [$\Lambda(\omega)$ and $\Omega(\omega)$ in the notation of
Refs. \cite{WKB1,WKB2,WKB3,Will}] are negligible. Substituting Eqs.
(\ref{Eq13}), (\ref{Eq14}), (\ref{Eq16}), and (\ref{Eq17}) into Eqs.
(29)-(31) of Ref. \cite{Will}, one indeed finds
$\Lambda=O(qQ/H(r_+))\ll 1$ and $\Omega=O((qQ/H(r_+))^2)\ll 1$ in
the $qQ\ll H(r_+)$ regime. Remembering that $H(r_+)\leq (l+1)^2$ and
taking cognizance of Eq. (\ref{Eq11}), one concludes that our
analytical WKB treatment is valid in the regime
%realizes that the QNM
%condition (\ref{Eq15}) is valid [that is, the higher-order
%correction terms satisfy $\Lambda(\omega)\ll 1$ and
%$\Omega(\omega)\ll 1$] provided $V^{(n)}/V^{(2)}\ll V^{(2)}$ at
%$y=y_0$ for $n\geq 3$, see \cite{WKB1,WKB2,WKB3,Will} for details.
%Taking cognizance of Eqs. (\ref{Eq13}), (\ref{Eq14}), (\ref{Eq16}),
%and (\ref{Eq17}), one finds that this condition is satisfied in the
%regime $H(r_+)\gg qQ$.
\begin{equation}\label{Eq18}
%1\ll
l\ll qQ \ll l^2\  .
\end{equation}
In this regime Eqs. (\ref{Eq16})-(\ref{Eq17}) can be simplified to
yield
\begin{equation}\label{Eq19}
\varpi_R={{H(r_+)(r_+-r_-)}\over{4q^2Q^2r_+}}\Big[1+O\Big({{q^2Q^2}\over{l^4}}\Big)\Big]
\end{equation}
and
\begin{equation}\label{Eq20}
\varpi_I=-i{{(r_+-r_-)}\over{2qQr_+}}(n+1/2)\Big[1+O\Big({{q^2Q^2}\over{l^4}}\Big)\Big]\
.
\end{equation}
Finally, taking cognizance of Eq. (\ref{Eq12}), one obtains
\begin{equation}\label{Eq21}
\omega_n={{qQ}\over{r_+}}+\pi T_{\text{BH}}{{l(l+1)}\over{qQ}}-i2\pi
T_{\text{BH}}\cdot (n+1/2)
\end{equation}
%and
%\begin{equation}\label{Eq5}
%|\omega_I|=2\pi T_{\text{BH}}\cdot (n+1/2)\  ,
%\end{equation}
for the quasinormal resonances of the charged scalar field in the
charged RN black-hole spacetime. Here
$T_{\text{BH}}={{r_+-r_-}\over{4\pi r^2_+}}$ is the
Bekenstein-Hawking temperature of the black hole.

%From Eq. (\ref{Eq}) one learns that
%\begin{equation}\label{Eq5}
%|\epsilon_I|\leq {{(r_+-r_-)(n+1/2)}\over{2qQr_+}}\  ,
%\end{equation}
%which implies [see Eq. (\ref{Eq})]
%\begin{equation}\label{Eq5}
%|\omega_I|\leq 2\pi T_{\text{BH}}(n+1/2)\  .
%\end{equation}

\section{Summary}

In summary, we have studied the spectrum of quasinormal resonances
which characterizes the dynamics of a charged scalar field in the
spacetime of a charged Reissner-Nordstr\"om black hole. We have
shown that this spectrum can be studied {\it analytically} in the
regime $l\ll qQ \ll l^2$. In particular, in this regime the
fundamental resonances can be expressed in terms of the black-hole
physical parameters: the electric potential $\Phi={{Q}\over{r_+}}$
and the temperature $T_{\text{BH}}$.

Finally, we recall that a universal bound on the relaxation time
$\tau_{\text{relax}}$ of a perturbed thermodynamical system has
recently been suggested
\cite{Hod1b,Hod2b,Gruzb,Pescib,Hodbb,Hod8b,Hod9b}:
\begin{equation}\label{Eq22}
\tau_{\text{relax}} \geq \hbar/\pi T\  ,
\end{equation}
where $T$ is the system's temperature. This bound can be regarded as
a quantitative formulation of the third law of thermodynamics. The
characteristic dynamical timescale for generic perturbations to
decay is given by $\tau=1/\omega^{\text{fund}}_I$, where
$\omega^{\text{fund}}$ is the fundamental (least damped) quasinormal
resonance of the black hole. The universal relaxation bound
(\ref{Eq22}) implies that the imaginary part of the fundamental
quasinormal resonance should conform to the upper bound
\begin{equation}\label{Eq23}
\omega^{\text{fund}}_I \leq \pi T_{BH}/\hbar\  .
\end{equation}
Taking cognizance of Eq. (\ref{Eq21}) for the quasinormal resonances
of the charged field and substituting $n=0$ for the fundamental
mode, one finds
\begin{equation}\label{Eq24}
\omega^{\text{fund}}_I=\pi T_{BH}/\hbar\  .
\end{equation}
We therefore conclude that the relaxation dynamics of perturbed RN
black holes may saturate the universal relaxation bound (\ref{Eq22})
in the $qQ\gg \hbar$ regime.

\bigskip
\noindent
{\bf ACKNOWLEDGMENTS}
\bigskip

This research is supported by the Carmel Science Foundation. We
thank Yael Oren, Arbel M. Ongo and Ayelet B. Lata for helpful
discussions.

%\newpage

\end{document}